# Feature Selection Parallel Technique for Remotely Sensed Imagery Classification

N-A. LeKhac, M-T. Kechadi
School of Computer Science & Informatics
University College Dublin
Dublin, Ireland
{an.lekhac,tahar.kechadi}@ucd.ie

B. Wu, C. Chen
Key Lab. of Spatial Data Mining and Information Sharing
Fuzhou University
Fuzhou, P.R. China
{b.wu,c.chen}@fzu.edu.cn

*Abstract*—Remote sensing research focusing on feature selection has long attracted the attention of the remote sensing community because feature selection is a prerequisite for image processing and various applications. Different feature selection methods have been proposed to improve the classification accuracy. They vary from basic search techniques to clonal selections, and various optimal criteria have been investigated. Recently, methods using dependence-based measures have attracted much attention due to their ability to deal with very high dimensional datasets. However, these methods are based on Cramer's V test, which has performance issues with large datasets. In this paper, we propose a parallel approach to improve their performance. We evaluate our approach on hyper-spectral and high spatial resolution images and compare it to the proposed methods with a centralized version as preliminary results. The results are very promising.

## I. Introduction

Remote sensing research that focuses on feature extraction and selection has attracted much attention of for quiet long time because feature extraction and selection constitute core prerequisites for various applications (e.g., image processing). Tremendous efforts have been dedicated to developing various feature extraction and selection methods to improve image processing effectiveness and classification accuracy in the last few decades [1][2][3]. Previous research generally suggests that effective extraction and utilization of potential multiple features of remotely sensed data, such as spectral signatures, various induced indices, and textural or contextual information, can significantly improve classification accuracy [4][5]. However, not all extracted features are equivalent in their contribution to classification tasks; some of them perhaps are superfluous and useless because either they are trivial or present high correlations between them. Accordingly, the use of all possible features in a classification procedure may add unnecessary information redundancy and significantly decrease the classification accuracy [6]. As a consequence, employing feature selection techniques to extract the most effective subset from the candidate features is critical to classification of remotely sensed data into a thematic map [5]. Given an input data with $N$ samples and $m$ features $X = \{x_1, ...,x_m\}$, and the target classification variable $c$, the problem is to find a subset $S$ of $n$ features from $m$ ($n \leq m$) features that optimally characterizes $c$.

Conceptually, feature selection in general requires a search strategy and criterion functions [7]. The search algorithm generates and compares possible feature selection solutions by calculating their criterion function values as a measure of the effectiveness of each given subset. Sequential search techniques, classical feature selection methods, look for the best feature subset with the prefixed number of features by adding to (resp. removing from) the current feature subset one feature at a time. These include sequential forward-selection (SFS) and sequential backward selection (SBS) [8]. Recently stochastic search algorithms, such as genetic algorithms [9] and clonal selection algorithms [10] have also been attempted for feature selection. Besides the search strategies, an optimal subset always depends of the evaluation function. Various optimal criteria, such as distance-based [11], entropy-based [12] and dependence-based measures [13], etc., have been widely investigated. In this paper, we focus on dependence based techniques. These techniques have some key advantages: 1) they can easily handle very high dimensional datasets (scalability), 2) they are computationally simple and fast, and 3) they are independent of the classification algorithm used. However, it has been recognized that a direct combination of individually good features in terms of certain criteria do not necessary lead to the best overall performance. As an alternative, some researchers have studied indirect or direct means to reduce redundancy among features and select features with minimal redundancy. [12] proposed a heuristic mutual information based max-dependency criterion (mRMR) method to minimize redundancy, which uses a series of intuitive measures of relevance to select very good features. The mRMR method was proven to be an effective technique for feature selection for remote sensing image classification[14]. Recently, we proposed two feature selection indices based on maximal association and minimal redundancy derived from *Cramer's V* test. We also showed its performance in terms of classification accuracy [15]. However, the main drawback of this method is the computational time. It is more severe in case of very large number of features. To analysis big-data of hundreds of features, we propose a parallel approach for feature selection. In order to evaluate this approach on large scale systems, we firstly show its performance with multi-threading paradigm and and MPI programming model.

The rest of the paper is organised as follows. In Section II we present the background of our research where we discuss feature selection, the max-min associated indices for feature selection, and parallel/distributed paradigm. Section III summarises briefly a centralisation-based approach for feature section, then we described in detail aparallel/distributed model for max-min associated indices for feature selection in Section IV. In Section V we evaluated the results of our model on sensored datasets and compared it to our previous approach. In Section VI we discuss future work and conclude.

## II. BACKGROUND

In this section, we start by briefly describing an approach of feature selection based on two associated indices, we then present models for parallel algorithms. Finally, we discuss some related work in the context of our approach.

### A. Indices for feature selection

*1) Cramer's V test:* The Chi-square test is one of the widely used measures to define dependence of variables and was proven to be effective in feature selection [13]. However, it is known that the Chi-square test of dependence is very sensitive to the sample size [16]. *Cramer's V* is the most popular nominal association that is used to measure the strength of the relationship between variables regardless of table size [17]. It has the advantage of not being affected by sample size and therefore is very useful in situations where one suspects a statistically significant Chi-square was the result of a large sample size rather than any substantive relationship between the variables [17]. Therefore, *Cramer's V* test is employed to measure the association between target and variables. Given a *r-row* by *s-column* cross-tabulation, *Cramer's V* can be directly derived from the Chi-square statistic:

$$V = \sqrt{\frac{\chi^2}{N \min\{(r-1),(s-1)\}}} \quad (1)$$

The value of *Cramer's V* varies between 0 and 1. If its value is large, then there is a tendency for particular categories of the first variable to be associated with particular categories of the second variable. It has been suggested in practice that a Cramer's V of 0.1 provides a good minimum threshold for suggesting there is a substantive relationship between two variables [18].

*2) Max-Min associated indices:* In [15], we explored the possibility that a combination of the Cramer's V coefficients can be further exploited for optimal feature selection. Two max-min-associated indices derived from the *Cramer's V* test coefficient were developed. For convenience, we firstly presented some important notations employed in this paper in Table 1.

TABLE I. NOTATIONS

| Notation | Description | Notation | Description |
|---|---|---|---|
| $X$ | Feature set | $C$ | Number of classes |
| $N$ | Total samples | $x_i$ | The $i^{th}$ feature |
| $m$ | Number of features | $N_k$ | Sample number of the kth cross-validation |
| $S$ | Selected subset | $V(x_i,c)$ | Cramer's V test between $x_i$ and target $c$ |
| $|S|$ | Number of element of set S | $V(x_i,x_j)$ | Cramer's V test between $x_i$ and $x_j$ |

Intuitively, selected features must have maximal target class associated ability. Therefore, a max-associated criterion is used to search for features satisfying (2) with *Cramer's V* test measurement between individual features $x_i$ and class $c$ (*A* condition):

$$\max A(S,c), A = \frac{1}{|S|} \sum_{x_i \in S} V(x_i, c) \quad (2)$$

where *S* is the number of subset features. It is likely that features selected according to the max-associated condition (2) will result in rich redundancy, i.e. the dependency among these features could be larger. When two features highly depend on each other, the respective class discriminated power would not change much if one of them was removed. Therefore, the following minimal associated condition (R condition) among selected features could be added to select mutually exclusive features:

$$\min R(S), R = \frac{1}{|S|^2} \sum_{x_i, x_j \in S} V(x_i, x_j) \quad (3)$$

The max-min-associated indices for feature selection are derived directly from the above two criteria. Two combined methods, referred to as MMAIQ and MMAIS, are designed. These combinations are expressed as Eq. (4) and (5).

$$max\phi(A, R), \phi=A/R \quad (4)$$

$$max\phi(A, R), \phi=A- \lambda R \quad (5)$$

It is apparent that both (4) and (5) simultaneously satisfy the constraints on A and R. That is, a good feature should be one with maximal target class associated ability, and at the same time with minimal association among the selected features. In Eq. (5), there is a regularization parameter *λ*, whose function is to balance the functions of the two constraints in (2) and (3).

### B. Parallel Computing models

Dividing the problem into smaller tasks and assigning them to different processors for parallel execution are the two key steps in the design of parallel algorithms. Normally, there are two parallel models: data-parallel and task-parallel model. The data-parallel model is one of the simplest. In this model, the tasks are statically or semi-statically mapped onto processors and each task performs similar operations on different data. This type of parallelism that is a result of identical operations being applied concurrently on different data items is called data parallelism. Since all processors perform similar computations, the decomposition of the problem into tasks is usually based on data partitioning because a uniform partitioning of data followed by a static mapping is sufficient to guarantee load balancing [19].

The computations in any parallel algorithm can be viewed as a task-dependency graph. The task-dependency graph may be either trivial, as in the case of matrix multiplication, or complicated. In the task graph model, the interrelationships among the tasks are utilized to promote locality or to reduce interaction costs. This model is typically employed to solve problems in which the amount of data associated with the tasks is large relative to the amount of computation associated with them. This type of parallelism that is naturally expressed by independent tasks in a task-dependency graph is called task parallelism [19].

### III. FEATURE SELECTION ALGORITHM

In [15], we presented a feature selection approach that can be summarised as follows: to select the candidate feature set, an incremental method is used to find the sub-optimal features defined by Eq. (4) or (5). Although this search strategy does not allow the features to be reselected once they have been selected, it can usually ensure that the selected features with relevance and redundancy constraints are the most prominent features not to be removed. In addition, the incremental search method is rather fast. Suppose we already have $S_{p-1}$, the set with $p^{th}$ features, the task is to select the $p^{th}$ feature from set $\{X - S_{p-1}\}$, such that the feature maximizes Eq. (4) or Eq. (5). The incremental algorithm optimizes the following conditions:

$$\max_{x_j \in X - S_{p-1}} [V(x_j, c) - \frac{\lambda}{p-1} \sum_{x_i, x_j \in S} V(x_j, x_i)] \quad (6)$$

$$\max_{x_j \in X - S_{p-1}} [V(x_j, c) / \frac{\lambda}{p-1} \sum_{x_i, x_j \in S} V(x_j, x_i)] \quad (7)$$

These optimizations can be computed efficiently in O(|S|·m) complexity. As a result, we can obtain the ranked features rapidly even if the dimension of features is possibly very high.

In {Bo}, two important issues were also solved before the classification process. One is how to obtain the cross-tabulation, such that *Cramer's V* can be calculated if the concerned features contain continuous variables. In this case, a pre-processing step of discretization is required to obtain cross-tabulation. Another critical problem is how to optimize the best number of feature subsets. The best number of features is usually estimated by the K folds cross-validation of the correct classification rate [15]. As shown in [15], this algorithm have been compared in terms of overall accuracy and kappa coefficient with the SFS and mRMR methods, which are known for their general abilities and good performances. When compared with SFS and mRMR, MMAIQ performs the best feature selection, and offers better or comparable classification accuracy in two experiments with different types of image. MMAIS also achieves satisfactory results in the same experiments. These results testify that MMAIQ and MMAIS provide new and effective options for feature selection.

Despite this approach is efficient in terms of accuracy in the feature selection, its running time is still an issues with large datasets. The overall complexity of this approach is $O(c^3 \times r^2 \times r\log(r))$ where $r$ is the number of samples and $c$ is the number of features and $r\log(r)$ is the complexity of sorting algorithm used in *genCT* function that creates a relevant cross-table. When the number of samples as well as the number of features is increased, the running time is exposed with exponential complexity. Therefore, we propose a parallel approach for this algorithm in this paper.

### IV. PARALLEL APPROACH

As mentioned in Section II.B, a parallel algorithm can be implemented in either data-parallelism or task parallelism. In our first approach, we apply the data-parallelism. The reason is that the generation of cross table as well as the calculation of Cramer's V Test require the analysis of whole datasets. Therefore, the overhead of exchanging large size of datasets among different sites should be taken into account in a data-parallel approach. As a consequence, we base on the task-parallel paradigm to design our parallel algorithm. We also presume that two function *genCT* (creating relevant cross-tables) and *CVTest* (Cramer's V Test) are atomic i.e. a parallel versions of *genCT* and *CVTest* are not in the context of this paper.

In analysing the sequential version of optimal Cramer's V [15], we notice that the cross validation step is an impact on runtime performance. The complexity of this algorithm is also based on this step as shown in the previous section. We propose hence, a parallel approach for this step. Algorithm 4.1 describes our approach. Let the input dataset be *DisX* that is a matrix of $r \times c$ where $r$ is the number of samples and $c$ is the number of features. Moreover, let $DisXc_i$ be the column $i$ of *DisX*, *Y* be the label set, *kk* be the number of repetitions, *genCT* be the function that generates a cross tabulation table and *CVTest* be the function that performs the *Cramer's V* test. The feature selection algorithm can be resumed as in the Algorithm 4.1 (pseudo code).

Moreover, there is a threshold $P_{THESHOLD}$ in this algorithm. The purpose of this threshold is to handle the granularity of this algorithm. When the number of iterations *left* is under this threshold, a sequential computation will be applied.

The function *incr(x)* in this algorithm is equivalent to $x \leftarrow x + 1$. In this algorithm, the function $Par_{Fold}()$ performs the parallel computation of cross-table different parts $DisX_k$ of input datasets on in $p$ processors. It can be described as in the algorithm $Par_{Fold}()$.

*Algorithm 4.1 (pseudo code) Parallel Optimal Cramer's V*

**Input:** *DisX, r, c, Y, kk.*

**Output:** List the feature selected *fea*

1: $crm \leftarrow \{0\}$;
2: **for** $i \leftarrow 1$ to $c$ **do**
3:     $crm_i \leftarrow CV\ Test(getGT\ (Y, r, DisXc_i))$;
4: **end for**
5: $last \leftarrow 1$; $fea_1 \leftarrow crm_1$; $curidx \leftarrow crm_2$;
6: $cmi \leftarrow \{0\}$; $tmi \leftarrow \{0\}$;
7: **for** $i \leftarrow 2$ to $kk$ **do**
8:     $fcln_i \leftarrow DisXc_{fealast}$;
9:     $left \leftarrow c - last$;
10:     **if** $left > P_{THRESHOLD}$ **then**
11:         **FORK** in k process $p_1, p_2, ..., p_k$
12:         **Each** process $p_k$ **do**
13:             $tbl_k \leftarrow ParFold(DisX_k, curidx, fcln_k)$;
14:         **endo**
15:         **JOIN** k process: $\{tbl\} = \cup\ tbl_k$
16:         **for** $j \leftarrow 0$ to $left$ **do**
17:             $cmi_{curidxj} \leftarrow cmi_{curidxj} + CV\ Test(tbl_i)$;
18:         **endfor**
19:     **else**
20:         **for** $j \leftarrow 1$ to $left$ **do**
21:             $tmi_j \leftarrow crm_j$;
22:             $tbl \leftarrow genCT(fcln_i, r, DisXc_{curidxj})$;
23:             $cmi_{curidxj} \leftarrow cmi_{curidxj} + CV\ Test(tbl)$;
24:         **endfor**
25:     **end if**
26:     $m \leftarrow i \times tmi_1 / cmi_{curidx1}$; $midx \leftarrow 1$;
27:     **for** $k \leftarrow 2$ to $left$ **do**
28:         $tmp \leftarrow i \times tmi_k / cmi_{curidxk}$;
29:         **if** $(tmp > m)$ **then**
30:             $m \leftarrow tmp$;
31:             $midx \leftarrow k$;
32:         **end if**
33:     **endfor**
34:     $fea_i \leftarrow curidx_{midx}$;
35:     $curidx_{midx} \leftarrow curidx$;
36:     $incr(last)$; $incr(curidx)$;
37: **endfor**
38: **for** $i \leftarrow 1$ to $kk$ **do**
39:     $incr(fea_i)$;
40: **end for**

---

*$Par_{Fold}(\ )$ function*

**Input:** *startPos, endPos, curidx.*

**Output:** *DisXk*

1: **for** $j \leftarrow startPos$ to $endPos$ **do**
2:     $cln \leftarrow GetMatrixColumn(curidx+j)$;
3:     $DisXk_j \leftarrow genCT(cln)$;
4: **end for**

## V. PERFORMANCE EVALUATION

### A. Experiments

In order to exploit efficiently different parallel computing platforms from shared-memory architecture to distributed architecture such as cluster and grid, we design two versions of our parallel approach. The first version is based on the multithreading paradigm where we can deploy on the shared-memory systems to exploit the multi-core architecture. This version is portable as it is based on the PTHREAD library [19]. The second version is based on an message-exchange paradigm to exploit the distributed architecture such as cluster or grid platforms. Concretely, it uses MPI library as the communication library because of it is simple and widely used as a standard of communication by exchanging message [19].

Besides, we uses different datasets to evaluate our approach. The first one is a hyperspectral remote sensing image (PHI) Xiaqiao PHI (*xq*). The data set used in this experiment was collected in September 1999 of the Xiaqiao test site, a mixed agricultural area in Changzhou city, Jiangsu province, China, and is airborne pushbroom hyperspectral imagery (PHI). A sub-scene (346Ã—350 pixels) of the PHI image with 80 bands was tested, and their spectral ranges were from 417 to 854 nm. Figure 1 shows the experimental PHI image cube. The ground truth spectral data were collected in September, 1999 by field spectrometer SE590. The observed image was expected to classify into eight representative classes, i.e. corn, vegetables-sweet potato, vegetable-cabbage, soil, float grass, road, water and puddle/polluted water. This dataset has 80 features. The second dataset is *fcl1* that is a historically significant data set, is located in the southern part of Tippecanoe County, Indiana. It has more than a few spectral bands, contains a significant number of vegetative species or ground cover classes, includes many regions (e.g., fields) containing a large numbers of contiguous pixels from a given class. This dataset has 12 features. The last one is *India* that is a district boundaries dataset prepared for FAO by Dept. of Energy and natural resources, University of Illinois, including

Coastlines, national-subnational boundaries, lakes, and Islands. This dataset has 185 features. Indeed, we test our approach on different platforms varied from multi-core (up to quad-core) to cluster computing architecture (up to four computational nodes).

TABLE II. PERFORMANCE OF PARALLEL OPTIMAL CRAMER'S V

| Datasets | Feature Selection Time (s) of OCV[a] | | |
|---|---|---|---|
| | *Sequential* | *PThread (2;4 threads)* | *MPI (2;4 nodes)* |
| *xq* | 25.53 | 15.77; 8.42 | 22.73; 19.92 |
| *flc1* | 0.37 | 0.28; 0.15 | 0.39; 0.41 |
| *India* | 293.18 | 168.73; 98.61 | 243.19; 139.81 |

[a.] Optimal Cramer's V

*B. Analysis*

As described in [15], the sequential optimal Cramer's V has been shown that it is the most efficient algorithm on Xiaqiao PHI dataset (*xq*) comparing to other selected algorithms with an improvement of 10.1% of overall accuracy. However, its computational time is significant. Table II compares the running time of two algorithms: sequential optimal Cramer's V and parallel optimal Cramer's V on different platforms with different datasets. By observing this table, we notice that the speed-up for multi-threading approach is around 60% i.e. we can improve the running time by 1.6 times. Moreover, the speed-up for cluster computing architecture depends on the testing dataset. It's 20% and 12% for the ***India*** and *xq* datasets respectively. However, we don't gain the speedup of fcl1 dataset. In this case, the communication time dominates the computational time. Analysis on the communication and computational time of parallel approaches can be found in [19]. This means our solution gets significant performance for big datasets (e.g. India) in terms of computational time.

Besides, experimental results also show that the running times of two functions *genCT* and *CVTest* take are insignificant compared to the overall computational time. It also proves our presumtion at the beginning of Section IV.

## VI. CONCLUSION AND FUTURE WORK

In this paper, we present a parallel approach for improving the performance of an optimal solution MMAIQ and MMAIS for the feature selection in the classification of remotely sensed imagery. We also evaluate our approach with different datasets and comparisons of the proposed methods with a centralisation version as preliminary results. These experimental results consistently show that the proposed methods can provide an effective solution for feature selection in improving significantly the computational time that is an drawback of most feature selection approaches.

A Hadoop/MapReduce version of this proposed method is implementing and testing to evaluate the robustness of our solution on large scale distributed systems of dozen to hundreds of computational nodes in the context of the feature selection for big data.

ACKNOWLEDGMENT

The authors gratefully acknowledge the research support from the EU Framework Programme 7, Marie Curie Actions under grant No. PIRSES-GA-2009-247608.